\documentclass[sigconf]{acmart}

\usepackage{algorithmic}
\usepackage{comment}
\usepackage{booktabs}
\usepackage{algorithm}
\usepackage{subcaption}
\usepackage{tikz}
\usetikzlibrary{quantikz2}

\AtBeginDocument{%
  }

\copyrightyear{2025}
\acmYear{2025}
\setcopyright{acmlicensed}\acmConference[GECCO '25]{Genetic and Evolutionary Computation Conference}{July 14--18, 2025}{Malaga, Spain}
\acmBooktitle{Genetic and Evolutionary Computation Conference (GECCO '25), July 14--18, 2025, Malaga, Spain}
\acmDOI{10.1145/3712256.3726392}
\acmISBN{979-8-4007-1465-8/2025/07}

\begin{document}

\title{Quantum Circuit Construction and Optimization through Hybrid Evolutionary Algorithms}


\author{Leo Sünkel}
\affiliation{%
  \institution{LMU Munich}
  \city{Munich}
  \country{Germany}}
\email{leo.suenkel@ifi.lmu.de}

\author{Philipp Altmann}
\affiliation{%
  \institution{LMU Munich}
  \city{Munich}
  \country{Germany}}

\author{Michael Kölle}
\affiliation{%
  \institution{LMU Munich}
  \city{Munich}
  \country{Germany}}

\author{Gerhard Stenzel}
\affiliation{%
  \institution{LMU Munich}
  \city{Munich}
  \country{Germany}}

\author{Thomas Gabor}
\affiliation{%
  \institution{LMU Munich}
  \city{Munich}
  \country{Germany}}

\author{Claudia Linnhoff-Popien}
\affiliation{%
  \institution{LMU Munich}
  \city{Munich}
  \country{Germany}}

\renewcommand{\shortauthors}{Sünkel et al.}

\begin{abstract}
We apply a hybrid evolutionary algorithm to minimize the depth of circuits in quantum computing. More specifically, we evaluate two different variants of the algorithm. In the first approach, we combine the evolutionary algorithm with an optimization subroutine to optimize the parameters of the rotation gates present in the quantum circuit. In the second, the algorithm solely relies on evolutionary operations (i.e., mutations and crossover). We approach the problem from two sides: (1) constructing circuits from the ground up by starting with random initializations and (2) initializing individuals with a target circuit in order to optimize it further according to the fitness function. We run experiments on random circuits with 4 and 6 qubits varying in circuit depth. Our results show that the proposed methods are able to significantly reduce the depth of circuits while still retaining a high fidelity to the target state.
\end{abstract}



\maketitle

\section{Introduction}
With increasing number of qubits in current quantum computers, the hope to apply these machines to solve larger, practically relevant problems also rises. As the field progresses further, many problems must first be overcome to achieve practical relevance when compared to classical machines. For one, quantum computers, or QPUs, must be scaled vastly further than currently possible. Additionally, noise and decoherence can cause inaccuracies in the computation, limiting the number of operations that can reliably be executed in a given circuit. While Shor's \cite{shor1994algorithms} and Grover's \cite{grover1996fast} algorithm provide an exponential and quadratic speedup respectively over classical counterparts, the quest for further algorithms providing the so-called quantum advantage is still ongoing. Moreover, designing quantum circuits to perform some task is quite different from writing classical programs, and designing efficient ones for different QPU architectures poses further challenges. Quantum circuit synthesis or quantum architecture search therefore are active fields of research, in which automated construction of circuits that perform some given task is the main objective. Evolutionary algorithms have been applied to a variety of problems in this domain \cite{lukac2003evolutionary,mukherjee2009synthesis,ruican2008genetic}; however, in recent years, other techniques, including reinforcement learning \cite{kremer2024practical,kolle2024reinforcement,ostaszewski2021reinforcement}, have also been studied.
The reduction of a circuit's depth is a crucial optimization step before running it on real hardware, as too many noisy gates may corrupt the computation. In this paper, we apply a hybrid multi-objective evolutionary algorithm (EA) to this problem. More specifically, the EA aims to construct quantum circuits with minimal depth while maximizing their fidelity to the given target state. We run experiments approaching the problem from two directions: (1) the EA constructs new circuits from scratch and (2) the EA starts with the given circuit and optimizes it in order to reduce its depth. The hybrid EA consists of a parameter optimization subroutine applied at different intervals. We compare this to an approach without this subroutine as well as two baselines. We show that the hybrid approach is able to drastically reduce a circuits depth while maintaining a high fidelity.

\section{Background}\label{sec:background}

\subsection{Quantum Computing}
We will limit our discussion of quantum computing to the necessary components required for this paper. For a more comprehensive introduction to the topic, we refer to \cite{nielsen2010quantum,watrous2018theory}. 

\subsubsection{Fundamentals}
The most fundamental building block in quantum computing is the qubit, which is defined as follows:

\begin{equation}
    |\psi\rangle = \alpha|0\rangle + \beta|1\rangle
\end{equation}

\noindent where $\alpha$ and $\beta$ are complex numbers representing probability amplitudes of the qubit being in either state (0 or 1) with |$\alpha$|² + |$\beta$|² = 1. Thus, a qubit can be in a linear combination or superposition of states, i.e., in multiple states simultaneously. However, once a qubit is measured, it collapses to one of its basis states, taking on a classical value. Multiple qubits can become entangled, a phenomenon in which their states become correlated even if the qubits are separated over large distances. This property has no counterpart in the classical world.
Operations on qubits can be performed by applying gates, which can be mathematically represented by unitary matrices. Below we list a few commonly used gates that are defined in the following way \cite{nielsen2010quantum}:

\begin{align*}
    I = \begin{bmatrix}
        1 & 0 \\
        0 & 1
    \end{bmatrix}
    && X = \begin{bmatrix}
        0 & 1 \\
        1 & 0
    \end{bmatrix} \\
    H = \frac{1}{\sqrt{2}} \begin{bmatrix}
    1 & 1 \\
    1 & -1
    \end{bmatrix} 
    && R_z(\theta) = \begin{bmatrix}
        e^{-i\theta/2} & 0 \\
        0 & e^{i\theta/2}
    \end{bmatrix}
\end{align*}

\noindent where, starting from the top left, we have the identity gate $I$, the $X$ or not gate, as well as the \emph{Hadamard} and the $RZ$ parameterized rotation gate on the bottom. In the following we have the $CX$ or controlled not gate, a multi-qubit gate that flips its target if the control qubit has the value 1.:

\begin{align*}
    CX = \begin{bmatrix}
        1 & 0 & 0 & 0 \\
        0 & 1 & 0 & 0 \\
        0 & 0 & 0 & 1 \\
        0 & 0 & 1 & 0
    \end{bmatrix}
\end{align*}

\begin{figure}[tb]
    \centering
    \begin{quantikz}
            \lstick{\ket{\psi_0}} & \gate{H} & \ctrl{1} & \gate{R_z(\theta_0)} & \meter{} \\
            \lstick{\ket{\psi_1}} & \gate{H} & \targ{} & \gate{X} & \meter{}
        \end{quantikz}
    \caption{Example of a random quantum circuit with 2 qubits and a depth of 3. The meter at the end symbolizes that the qubits are being measured.}
    \label{fig:example_qc}
\end{figure}
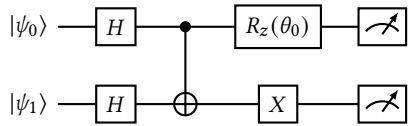

Having introduced qubits and gates, we can now define a quantum circuit (QC). In a QC, a horizontal line represents a single qubit, and each operation performed on the qubit is specified by a gate. Time flows from left to right, i.e., the first gate performed is the one on the far left. An example QC is given in Fig. \ref{fig:example_qc}. In this example, the circuit contains two qubits with 3 gates each and a measurement operation and the end.

An important concept from quantum information theory is the notion of fidelity, which  is a measure on how close two quantum states are. 
For two quantum states $\rho$ and $\sigma$ it is defined as \cite{nielsen2010quantum}:

\begin{equation}
    F(\rho, \sigma) = tr\sqrt{\sqrt{\rho} \sigma \sqrt{\rho}}
\end{equation}

For a pure state $\psi$ and an arbitrary state $\rho$ it is defined as \cite{nielsen2010quantum}:

\begin{equation}
    F(|\psi\rangle, \rho) = \sqrt{\langle\psi|\rho|\psi\rangle}
\end{equation}

\noindent where $\psi$ and $\rho$ are quantum states and a fidelity of 1 signifies identical and 0 orthogonal states.

\subsubsection{Current Era}
Due to the nature of quantum physics, qubits are inherently fragile and must be protected from external influence to prevent decoherence. Furthermore, gates are noisy operations and introduce errors into the computation. These are some of the key problems in the current state of quantum computing \cite{preskill2018quantum,alvarez2024simulating,hegde2024beyond,senapati2023towards}. It follows that circuits must be kept minimal, i.e., the depth of a circuit and the number of total gates that reliably can be executed is limited, and as the states decohere over time, the total execution time is also constrained.
Thus, the optimization of circuits becomes a crucial task for practical applications on real hardware, which is the topic of this work. 

\subsection{Evolutionary Algorithms}
\begin{figure}[tb]
    \centering
    \includegraphics[scale=0.3]{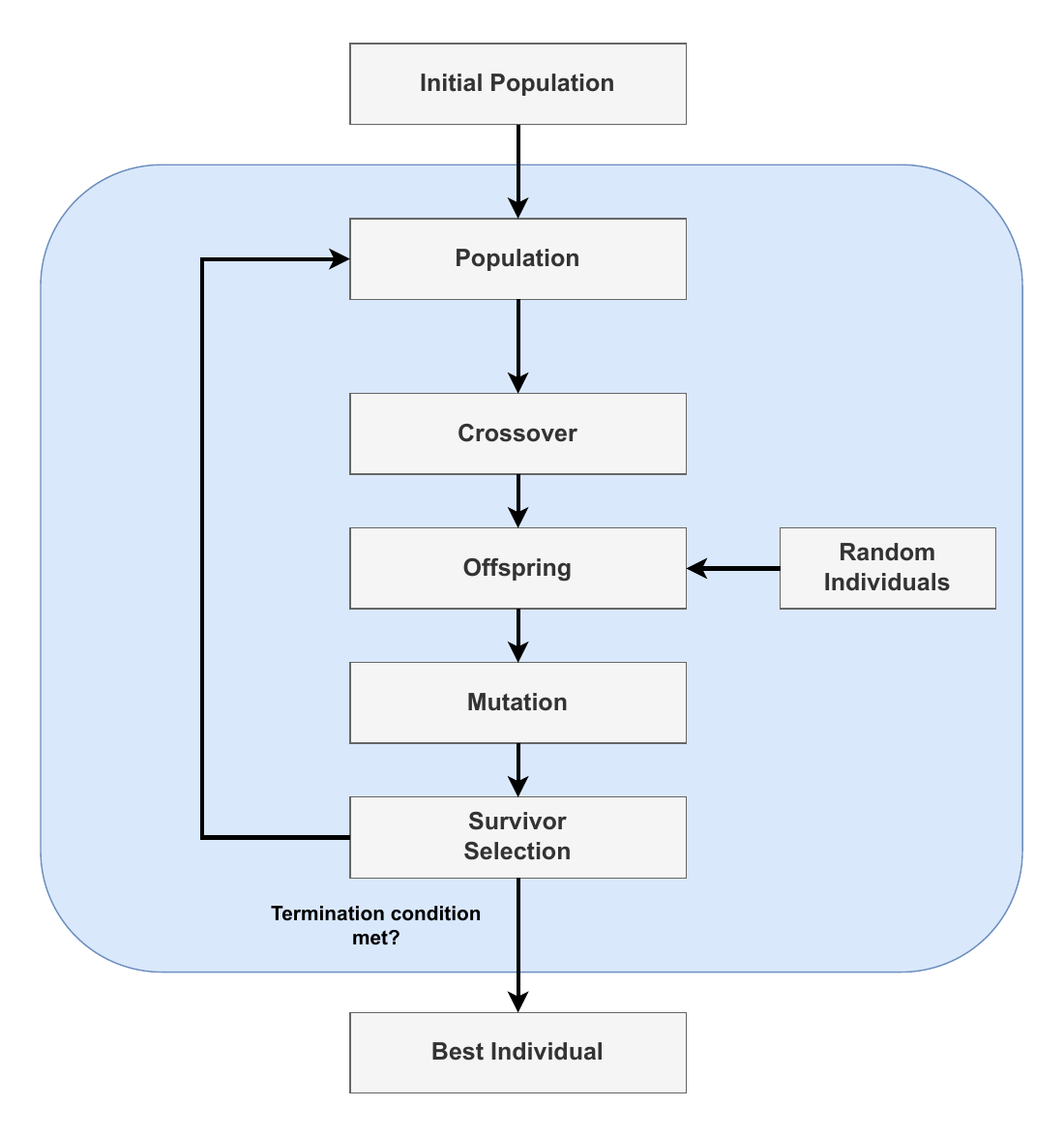}
    \caption{High level overview of the general process of an evolutionary algorithm.}
    \label{fig:ga_overview}
\end{figure}

An EA is a technique to solve optimization problems by applying methods commonly associated with processes found in biological evolution \cite{holland1992adaptation,eiben2002evolutionary}. It is a population-based approach where individual solutions are created by combining parts of other solutions and are subject to further modification according to random factors. The act of combining parts of existing solutions is known as \textit{crossover}, whereby a new individual is created through inheriting characteristics, i.e., genes, from two parent individuals. The hereby created solution can be referred to as the child or offspring. The offspring is then subjected to random mutations, i.e., random changes to its genes. In each \textit{generation} of the algorithm, new offspring are created and sorted by some \textit{fitness function}. The fitness value determines which individuals are transferred to the next generation and which are not. Thus, the main idea behind an EA is to gradually combine and improve existing solutions, thereby balancing exploitation and exploration and ultimately maximizing the fitness value of the objective function. An overview of this process is shown in Fig. \ref{fig:ga_overview}.

\subsection{Quantum Circuit Architecture Design}
Designing the architecture of a quantum circuit is not always a straightforward task, particularly when it is to be optimized under some constraints, for instance, due to hardware restrictions in specific devices or application related limitations. For quantum machine learning (QML) or variational quantum algorithms (VQAs) \cite{cerezo2021variational} in general, existing templates can commonly be utilized. However, what circuit is best suitable often remains elusive and is dependent on the task and other ramifications (e.g., what hardware is available). Automating this process has been increasingly becoming a subject of study in the quantum computing research community recently; particularly methods based on EAs or reinforcement learning (RL) are popular contestants in this field.
When we use the terms circuit optimization or quantum circuit architecture design in this paper, we essentially refer to two different approaches: (1) searching, i.e., building a quantum circuit from scratch that fulfills some constraints, and (2) optimizing an existing circuit under given constraints. A use case for the first approach could be to find novel circuit designs that prepare some quantum state, e.g., given a target state, find a circuit that prepares a state with maximal fidelity to the target. 
In the second approach, one already has an existing circuit and wants to optimize it, for instance, reduce the circuits depth or number of multi-qubit gates, or optimize it with respect to a particular quantum device. 
We thus define the task of automatic circuit design as applying an algorithm to search for quantum circuits that (1) fulfill a desired task and (2) adhere to the given constraints. 

\section{Related Work}\label{sec:related_work}
In \cite{potovcek2018multi}, the authors propose a multi-objective genetic algorithm for circuit construction. A hybrid approach to design quantum feature maps is discussed in \cite{pellow2024hybrid} whereas a multi-objective genetic algorithm for this task in \cite{altares2021automatic}. An EA was applied to quantum architecture search in \cite{zhang2023evolutionary}. A genetic algorithm is applied to quantum circuit compilation in \cite{rasconi2019innovative}. In \cite{chen2024evolutionary}, the authors apply an evolutionary approach to design variational quantum circuits. Evolutionary approaches to quantum architecture search and synthesis are also discussed in \cite{ding2022evolutionary} and \cite{lukac2003evolutionary}. 
RL is another popular approach and has been applied to a wide range of tasks in the domain of automated circuit construction. In \cite{ostaszewski2021reinforcement} the authors apply RL to optimize VQC architectures or QML models \cite{dai2024quantum}. Circuit or unitary synthesis is discussed in \cite{kolle2024reinforcement,kremer2024practical, rietsch2024unitary,Altmann24-C4QCD} while RL for compiling in \cite{quetschlich2023compiler}.
Other approaches including diffusion models \cite{furrutter2024quantum} or machine learning \cite{arrazola2019machine} have also been proposed. An overview of techniques for quantum circuit synthesis is given in \cite{ge2024quantum}.

\section{Approach}\label{sec:approach}
We introduce our approach in this section. We start with the problem, i.e., circuit encoding and define the individuals of the population. We then discuss the evolutionary operations and selection mechanisms. Finally, we discuss the two optimization subroutines before concluding the section with the definition of the fitness function.

\subsection{Circuit Encoding}
Circuits are encoded using two-dimensional lists, i.e., a matrix in which each row corresponds to a qubit and a column to a point in time of a particular operation. Each cell contains a gate object consisting of all necessary information (e.g., name, parameter, target, and control qubits). We refer to this representation as the solution of a particular individual. This solution representation was also utilized in \cite{sunkel2023ga4qco}, and the EA in this work is inspired by techniques introduced there. Each solution is converted to a Qiskit \cite{qiskit2024} quantum circuit so that it can be executed and applied to a particular task, and modifications to the circuit during the execution of the EA are applied on the solution. An example solution is shown in Table \ref{tab:example_solution} where each gate is represented by its name. 

\begin{table}[tb]
    \caption{Example solution representation of a circuit with 4 qubits and a depth of 4.}
    \label{tab:example_solution}
    \centering
    \begin{tabular}{|c|c|c|c|}
    \hline
    H & CX & H & SX \\ \hline 
    ID & H & CX & X  \\ \hline
    CX & CX & CX & ID \\ \hline
    CX & H & SX & ID \\ \hline
    \end{tabular}
\end{table}

\subsection{Individual}
The main components of an individual are the solution, a corresponding circuit implemented in Qiskit, and a fitness value. Each individual is initialized with a random solution, i.e., a random depth and random gates selected from the gate set specified below. Alternatively, a solution can be initialized with the target in the case where an existing circuit should be optimized. Available gates are specified by a basis gate set, and in principle the algorithm supports any gate set; however, in this work we only employ the sets shown in Table \ref{tab:ea_parameters}. Random gates can be single, multi or parameterized gates. Parameters are selected from the range $[0, 2\pi]$. If a multi-qubit (CX gate) is selected for a particular qubit, the procedure searches for another qubit available (i.e., not yet assigned at that time step) qubit. If an available qubit has been found, it is randomly determined which qubit is the control and which the target. If no available qubit is found, the qubit is assigned a random single qubit gate, ensuring that each qubit is assigned a gate at each time step. However, the gate sets used also contain identity gates, avoiding enforcing computations for all qubits.

\subsection{Crossover and Mutation}
All crossover and mutation methods implemented are introduced next. We will first describe crossover and how new solutions are created before describing the various available mutations.

\subsubsection{Crossover} 
\begin{figure*}[t]
    \centering
    \scalebox{0.8}{
        \begin{subfigure}[t]{0.45\textwidth}
            \centering
            \begin{quantikz}
                    \lstick{\ket{\psi_0}} & \gate{H} & \gate{X} & \gate{ID} & \gate{SX}  \\
                    \lstick{\ket{\psi_1}} & \gate{R_{x(\theta)}} &  \gate{SX} & \gate{H} & \gate{X} 
            \end{quantikz}
            \caption{Parent 1}
        \end{subfigure}
        \begin{subfigure}[t]{0.45\textwidth}
            \centering
            \begin{quantikz}
                    \lstick{\ket{\psi_0}} & \ctrl{1} & \gate{R_{x(\theta)}} & \ctrl{1}  & \gate{H}  \\
                    \lstick{\ket{\psi_1}} & \targ{} &  \gate{SX} & \targ{} & \gate{H} 
            \end{quantikz}
             \caption{Parent 2}
        \end{subfigure}
    }
    \scalebox{0.8}{
        \begin{subfigure}[t]{0.45\textwidth}
            \centering
            \begin{quantikz}
                \lstick{\ket{\psi_0}} & \gate{H} \gategroup[wires=2, steps=2, style={dashed, rounded corners, fill=red!20}, background]{} & \gate{X} & \ctrl{1} \gategroup[wires=2, steps=2, style={dashed, rounded corners, fill=blue!20}, background]{} & \gate{H}  \\
                \lstick{\ket{\psi_1}}& \gate{R_{x(\theta)}} &  \gate{SX} & \targ{} & \gate{H} 
            \end{quantikz}
            \caption{Child created with single point crossover where the cutoff point is after the second layer. Red is taken from parent 1 and blue from parent 2.}
            \label{fig:single_point_crossover}
        \end{subfigure}
        \begin{subfigure}[t]{0.45\textwidth}
            \centering
            \begin{quantikz}
                \lstick{\ket{\psi_0}} & \ctrl{1} \gategroup[wires=2, steps=1, style={dashed, rounded corners, fill=blue!20}, background]{} & \gate{X} \gategroup[wires=2, steps=1, style={dashed, rounded corners, fill=red!20}, background]{} & \ctrl{1} \gategroup[wires=2, steps=2, style={dashed, rounded corners, fill=blue!20}, background]{}  & \gate{H}  \\
                \lstick{\ket{\psi_1}} & \targ{} &  \gate{SX} & \targ{} & \gate{H} 
            \end{quantikz}
            \caption{Child created through uniform crossover where each column is selected from either parent randomly. Red is from parent 1 and blue parent 2.}
            \label{fig:uniform_crossover}
        \end{subfigure}
    }
    \caption{Overview of the crossover methods utilized by the EA.}
    \label{fig:crossover_overview}
\end{figure*}
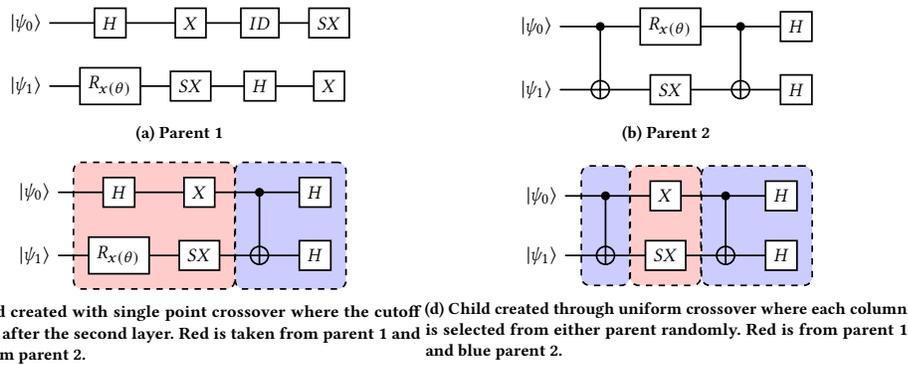

The EA makes use of two different crossover functions; which one is used is determined randomly for each child. Additionally, children can be created entirely random or inherit their solution from a single parent. Whether crossover is performed is determined randomly (cf. crossover rate in Tab. \ref{tab:ea_parameters}). If no crossover is performed, the child is created by either method mentioned above; which method is used is determined randomly.  
\textbf{Single Point Crossover}
As described above, the single point crossover randomly chooses a position that determines what part of the solution is inherited from either parent. As we allow for parents to have a different solution size, i.e., circuit depth, a slight modification is required. In addition to determining the crossover point randomly, the child inherits its circuit depth randomly from either parent. In the case where a child is larger than one parent, it inherits its gates from the smaller parent until the crossover point, and the rest from the larger parent. This process is illustrated in Fig. \ref{fig:single_point_crossover}.
\textbf{Uniform Column Crossover} 
In uniform crossover method, a child inherits each column of the solution matrix from either parent with an equal probability. As with the single point crossover, the child inherits its size from either parent. In the case where the child inherits a larger solution of one parent, the missing gates are filled up with gates from the larger parent. This crossover method is depicted in Fig. \ref{fig:uniform_crossover}.

\subsubsection{Mutation}
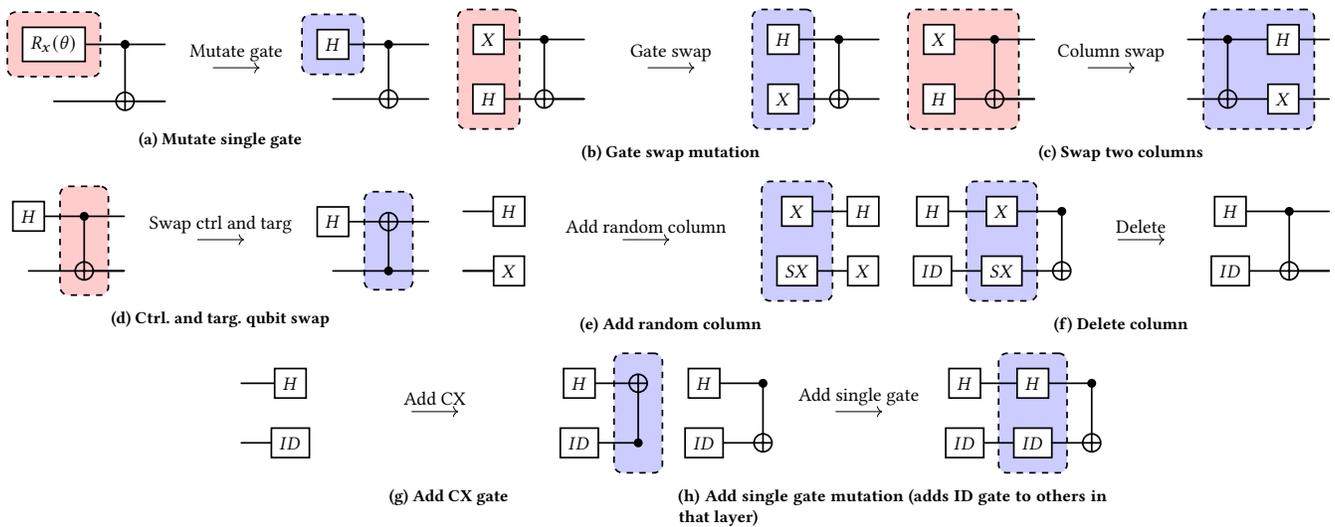
\begin{figure*}
    \scalebox{0.8}{
        \begin{subfigure}[t]{0.4\textwidth}
            \centering
            \begin{quantikz}
                \gate{R_x{(\theta)}} \gategroup[wires=1, steps=1, style={dashed, rounded corners, fill=red!20}, background]{} & \ctrl{1} & \\
                \qw & \targ{} & \qw
            \end{quantikz}
            \hfill
            \begin{tikzpicture}
                \node[] (A) at (0, 0) {};
                \node[] (B) at (1, 0) {} ;
                \draw[->] (A) -- node[above]{Mutate gate} (B);
            \end{tikzpicture}
            \hfill
            \begin{quantikz}
                \gate{H} \gategroup[wires=1, steps=1, style={dashed, rounded corners, fill=blue!20}, background]{} & \ctrl{1} & \\
                & \targ{} &
            \end{quantikz}
        \caption{Mutate single gate}
         \end{subfigure}
     }
     \hfill
     \scalebox{0.8}{
         \begin{subfigure}[t]{0.4\textwidth}
            \centering
            \begin{quantikz}
                \gate{X} \gategroup[wires=2, steps=1, style={dashed, rounded corners, fill=red!20}, background]{} & \ctrl{1} & \\
                \gate{H} \qw & \targ{} & \qw
            \end{quantikz}
            \hfill
            \begin{tikzpicture}
                \node[] (A) at (0, 0) {};
                \node[] (B) at (1, 0) {} ;
                \draw[->] (A) -- node[above]{Gate swap} (B);
            \end{tikzpicture}
            \hfill
            \begin{quantikz}
                \gate{H} \gategroup[wires=2, steps=1, style={dashed, rounded corners, fill=blue!20}, background]{} & \ctrl{1} & \\
               \gate{X} & \targ{} &
            \end{quantikz}
        \caption{Gate swap mutation}
        \end{subfigure}
    }
    \hfill
    \scalebox{0.8}{
        \begin{subfigure}[t]{0.4\textwidth}
            \begin{quantikz}
                \gate{X} \gategroup[wires=2, steps=2, style={dashed, rounded corners, fill=red!20}, background]{} & \ctrl{1} & \\
                \gate{H} \qw & \targ{} & \qw
            \end{quantikz}
            \hfill
            \begin{tikzpicture}
                \node[] (A) at (0, 0) {};
                \node[] (B) at (1, 0) {} ;
                \draw[->] (A) -- node[above]{Column swap} (B);
            \end{tikzpicture}
            \hfill
            \begin{quantikz}
                 & \ctrl{1}  \gategroup[wires=2, steps=2, style={dashed, rounded corners, fill=blue!20}, background]{}& \gate{H}   & \\
                 & \targ{} & \gate{X} & 
            \end{quantikz}
        \caption{Swap two columns}
        \end{subfigure}
    }
    \hfill
    \scalebox{0.8}{
        \begin{subfigure}[t]{0.4\textwidth}
               \begin{quantikz}
                     \gate{H}  & \ctrl{1} \gategroup[wires=2, steps=1, style={dashed, rounded corners, fill=red!20}, background]{}& \\
                     \qw & \targ{} & \qw
                \end{quantikz}
                \hfill
                \begin{tikzpicture}
                    \node[] (A) at (0, 0) {};
                    \node[] (B) at (1, 0) {} ;
                    \draw[->] (A) -- node[above]{Swap ctrl and targ} (B);
                \end{tikzpicture}
                \hfill
                \begin{quantikz}
                     \gate{H}  & \targ{0} \gategroup[wires=2, steps=1, style={dashed, rounded corners, fill=blue!20}, background]{} & \\
                    & \ctrl{-1} &
                \end{quantikz}
        \caption{Ctrl. and targ. qubit swap}
        \end{subfigure}
    }
    \hfill
    \scalebox{0.8}{
        \begin{subfigure}[t]{0.4\textwidth}
            \begin{quantikz}
                & \gate{H}  \\
                & \gate{X}  \qw
            \end{quantikz}
            \hfill
            \begin{tikzpicture}
                \node[] (A) at (0, 0) {};
                \node[] (B) at (1, 0) {} ;
                \draw[->] (A) -- node[above]{Add random column} (B);
            \end{tikzpicture}
            \hfill
            \begin{quantikz}
                \gate{X} \gategroup[wires=2, steps=1, style={dashed, rounded corners, fill=blue!20}, background]{} & \gate{H}   \\
                \gate{SX} & \gate{X} 
            \end{quantikz}
        \caption{Add random column}
        \end{subfigure}
    }
    \hfill
    \scalebox{0.8}{
        \begin{subfigure}[t]{0.4\textwidth}
            \begin{quantikz}
                \gate{H} & \gate{X} \gategroup[wires=2, steps=1, style={dashed, rounded corners, fill=blue!20}, background]{} & \ctrl{1}   \\
                \gate{ID} & \gate{SX} & \targ{0} 
            \end{quantikz}
            \hfill
            \begin{tikzpicture}
                \node[] (A) at (0, 0) {};
                \node[] (B) at (1, 0) {} ;
                \draw[->] (A) -- node[above]{Delete} (B);
            \end{tikzpicture}
            \hfill
            \begin{quantikz}
                \gate{H}  & \ctrl{1} & \\
                \gate{ID} & \targ{} & \qw
            \end{quantikz}
        \caption{Delete column}
        \end{subfigure}
    }
    \hfill
    \scalebox{0.8}{
        \begin{subfigure}[t]{0.4\textwidth}
            \begin{quantikz}
                & \gate{H}   \\
                & \gate{ID} 
            \end{quantikz}
            \hfill
            \begin{tikzpicture}
                \node[] (A) at (0, 0) {};
                \node[] (B) at (1, 0) {} ;
                \draw[->] (A) -- node[above]{Add CX} (B);
            \end{tikzpicture}
            \hfill
            \begin{quantikz}
                \gate{H} & \targ{} \gategroup[wires=2, steps=1, style={dashed, rounded corners, fill=blue!20}, background]{}    \\
                \gate{ID} & \ctrl{-1} 
            \end{quantikz}
        \caption{Add CX gate}
        \end{subfigure}
    }
    \scalebox{0.8}{
    \begin{subfigure}[t]{0.4\textwidth}
            \begin{quantikz}
                \gate{H} & \ctrl{1}    \\
                \gate{ID} & \targ{0} 
            \end{quantikz}
            \hfill
            \begin{tikzpicture}
                \node[] (A) at (0, 0) {};
                \node[] (B) at (1, 0) {} ;
                \draw[->] (A) -- node[above]{Add single gate} (B);
            \end{tikzpicture}
            \hfill
            \begin{quantikz}
                \gate{H} & \gate{H} \gategroup[wires=2, steps=1, style={dashed, rounded corners, fill=blue!20}, background]{} & \ctrl{1}   \\
                \gate{ID} & \gate{ID} & \targ{0} 
            \end{quantikz}
        \caption{Add single gate mutation (adds ID gate to others in that layer)}
        \end{subfigure}
    }
    \caption{Overview of the mutation operations the algorithm utilizes. What method is employed is determined randomly.}
    \label{fig:mutation_overview}
\end{figure*}

The EA employs a wide range of mutation functions, and which one is selected is determined randomly. That is, after a child has been created according to the principles discussed in the previous section, the \textit{mutation rate} (cf. Tab. \ref{tab:ea_parameters}) determines if a mutation should be applied. Each method is visualized in Fig. \ref{fig:mutation_overview}. Note that in the case where new gates are added, i.e., a new layer (e.g., when adding a single qubit gate or a CX gate), the qubits not involved add an identity gate so that the solution matrix remains consistent. Also, if a CX gate is mutated by the ``mutate gate'' operation, both qubits, i.e., control and target, are mutated to single qubit gates. Gate swap operates on single qubit gates while ``swap ctrl and targ'' on CX gates.

\subsection{Selection}
Selection is relevant at two different stages: parent selection and survivor selection. In the former, the individuals that are combined through crossover are selected, whereas the latter determines what individuals are allowed to continue into the next generation. 
In the experiments part of this work, all individuals have an equal probability to be chosen as parents irrespective of fitness value. However, selection pressure is applied during survivor selection: here the worst $n$ individuals are replaced by the best $n$ children. More specifically, first $m$ children are created and sorted by their fitness; from this list the best $n$ are selected to replace the worst individuals in the population. The number of children to create in each generation is determined by the \textit{offspring rate} parameter where the number of individuals to replace in the population by the \textit{replace rate} parameter, cf. Table \ref{tab:ea_parameters}.

\subsection{Parameter Optimization}
In the hybrid EA, parameters of all parameterized gates present in a circuit, i.e., the rotation angles, are optimized every $n$ generations (cf. Table \ref{tab:ea_parameters}) of individuals in a random subset of the population. For this, the COBYLA optimizer provided by scikit-kit learn \cite{scikit-learn} is applied to minimize the following objective function:

\begin{equation}
    \min_\theta f(U, \theta, \rho) = 1 - F(U(\theta), \rho)
\end{equation}

\noindent with $U$ being the quantum circuit prepared by an individual and $\rho$ the target state. This subroutine is used to optimize the parameters in such a way to maximize the fidelity of the resulting state to the target, and thus the circuit itself is not changed. The initial parameters are set by each individual. Note that not every individual is being optimized, but rather a subset (0.1 of the population size) is randomly selected; this is done every 25 generations.

\subsection{Circuit Optimization}
Additionally to the parameter optimization subroutine, our approach also uses a subroutine to optimize the solutions itself. However, this subroutine does not change the functionality of the corresponding circuit; it rather optimizes its representation, i.e., the solutions are made more compact by combining rotations, moving gates forward when possible, and removing layers that only contain identity gates. However, this subroutine is not complete or exhaustive, i.e., further optimization may be possible; the aim here is to simply provide a quick heuristic to simplify individuals in the population.

\subsection{Fitness Function}
The objective in our experiments consists of two parts, i.e., we consider multi-objective EAs: (1) maximize the fidelity between the individual and target state (2) minimize the depth of the corresponding circuit. We define the components as follows:

\begin{equation}
   A = 1 - \delta_{norm}
\end{equation}
where $\delta$ is the normalized depth of the circuit. That is:

\begin{equation}
    \delta_{norm} = \frac{\delta - 1}{d - 1}
\end{equation}
where d is the depth of the target circuit and 1 the minimum depth a circuit can have. We assume $d>1$ as a target circuit with depth 1 cannot be optimized in terms of depth any further.

The second component is defined as:

\begin{equation}
    B = 1 - F(U, \rho)
\end{equation}
where $U$ is a quantum circuit corresponding to a solution of an individual and $\rho$ is the target state.
The fitness becomes:

\begin{align}
    \max f(U, \rho) = A - B \notag \\ 
    = F(U, \rho) - \delta_{norm} 
\end{align}

By introducing weights the fitness becomes:

\begin{equation}
    \max f(U, \rho) = \alpha * F(U, \rho) - \beta * \delta_{norm}
\end{equation}

In the experiments $\alpha$ was set to 10 and $\beta$ to 1. We use Qiskit to calculate the fidelity F.

\section{Experimental Setup}\label{sec:experimental_setup}
We apply the algorithm in two different ways: In the first, circuits are created from scratch, i.e., initial solutions are created randomly and the EA searches for circuits that maximize the fitness. In the second, individuals are initialized as the target circuit and the aim is to optimize an existing circuit while maximizing the same fitness as above. 
The experiments were performed on various randomly created circuits using different variants of the EA introduced above. The first approach is the hybrid EA, which is then compared to (1) the same EA with no parameter optimization subroutine, i.e., a regular EA, and (2) the hybrid approach with all EA operations (i.e., mutation and crossover) turned off. The purpose of the latter experiments is to verify that the EA operations add value and the parameter optimization is not solely responsible for the performance, and can be considered the first baseline. We also run experiments for a random baseline. Here in each generation a number equal to the number of children in the EA is created randomly with no further modification or optimization. Experiments were run for 4 different seeds and the plots show the mean of the aggregated results. The hyperparameters used are shown in Table \ref{tab:ea_parameters}, which were determined experimentally. 

\begin{table}[tb]
    \centering
    \caption{Hyperparameters of the EA and gateset used in circuits.}
    \label{tab:ea_parameters}
    \begin{tabular}{lr}
        \toprule
         Population size & 200  \\ \midrule
         Generations & 1000 \\ \midrule
         Crossover rate & 0.85 \\ \midrule
         Mutation rate & 0.85 \\ \midrule
         Offspring rate & 0.3 \\ \midrule
         Replace rate & 0.3 \\ \midrule
         Gate set & ID, X, SX, RZ and CX \\ \midrule
         Max COBYLA iterations & 1000 \\
        \bottomrule
    \end{tabular}
\end{table}

\begin{figure*}[t]
    \centering
    \begin{subfigure}[t]{0.45\textwidth}
        \centering
        \includegraphics[scale=0.3]{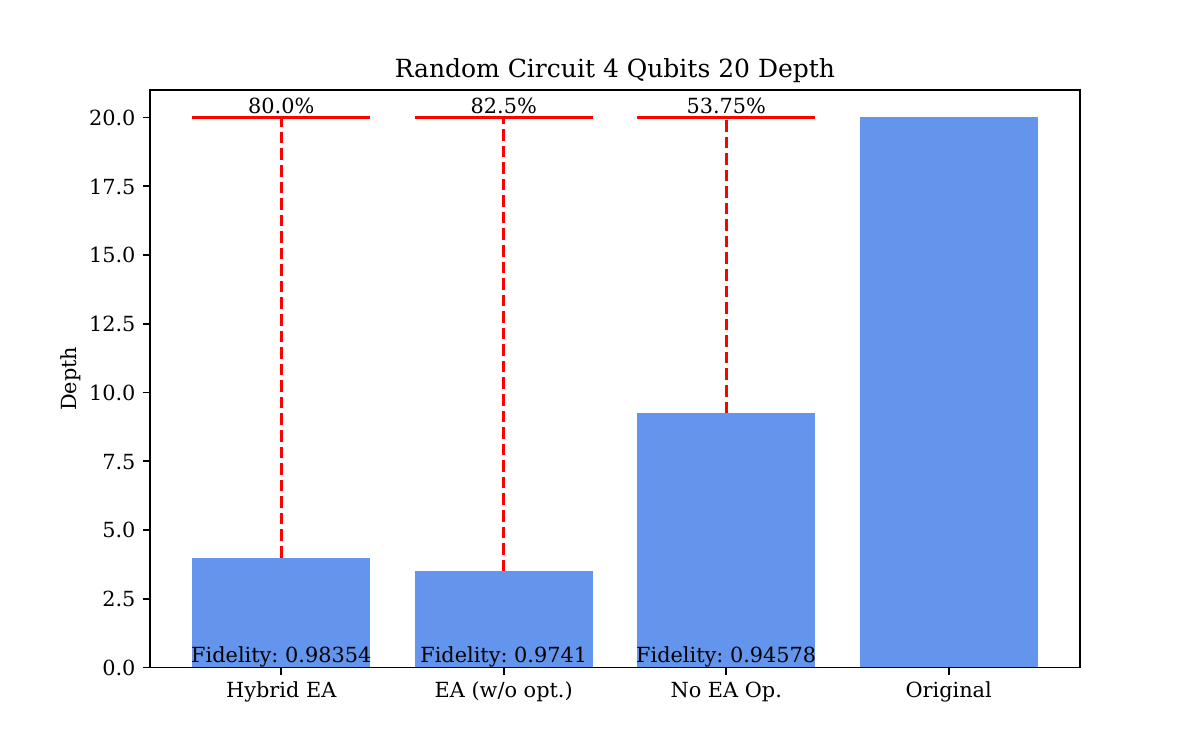}
        \caption{4 qubits and 20 depth.}
        \label{fig:results_qc_from_scratch_a}
    \end{subfigure}
    \begin{subfigure}[t]{0.45\textwidth}
        \centering
        \includegraphics[scale=0.3]{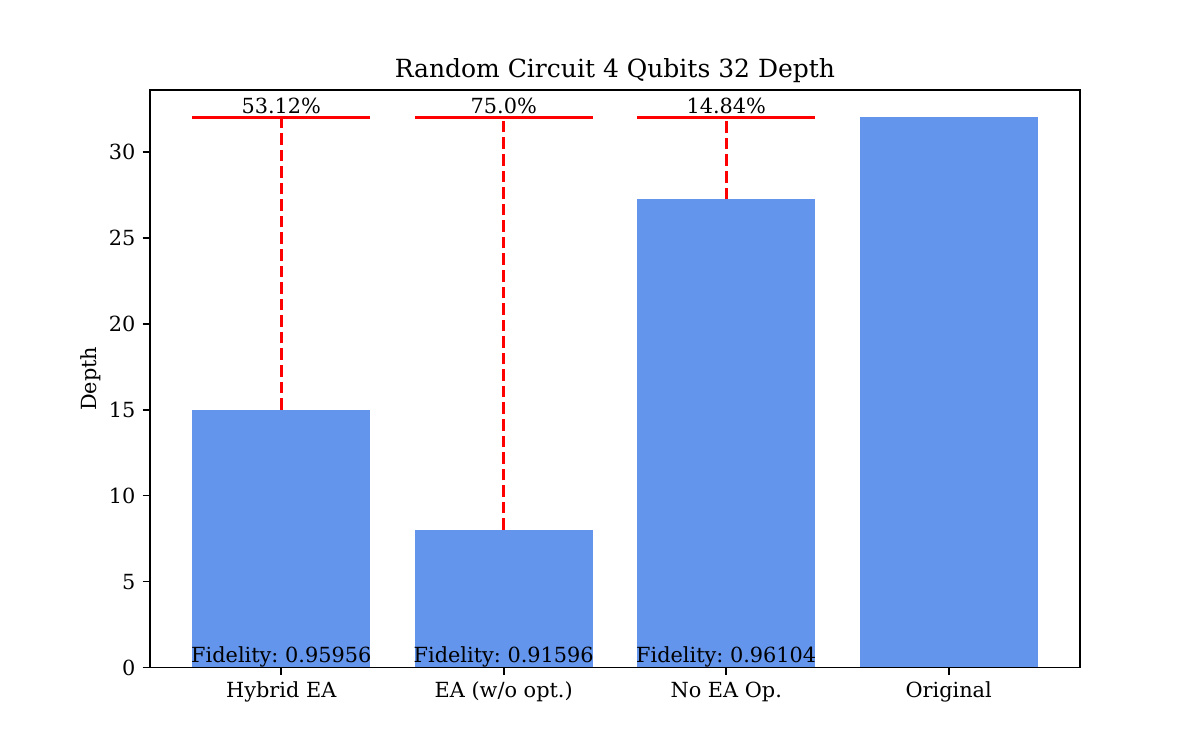}
        \caption{4 qubits and 32 depth.}
        \label{fig:results_qc_from_scratch_b}
    \end{subfigure}
    \begin{subfigure}[t]{0.45\textwidth}
        \centering
        \includegraphics[scale=0.3]{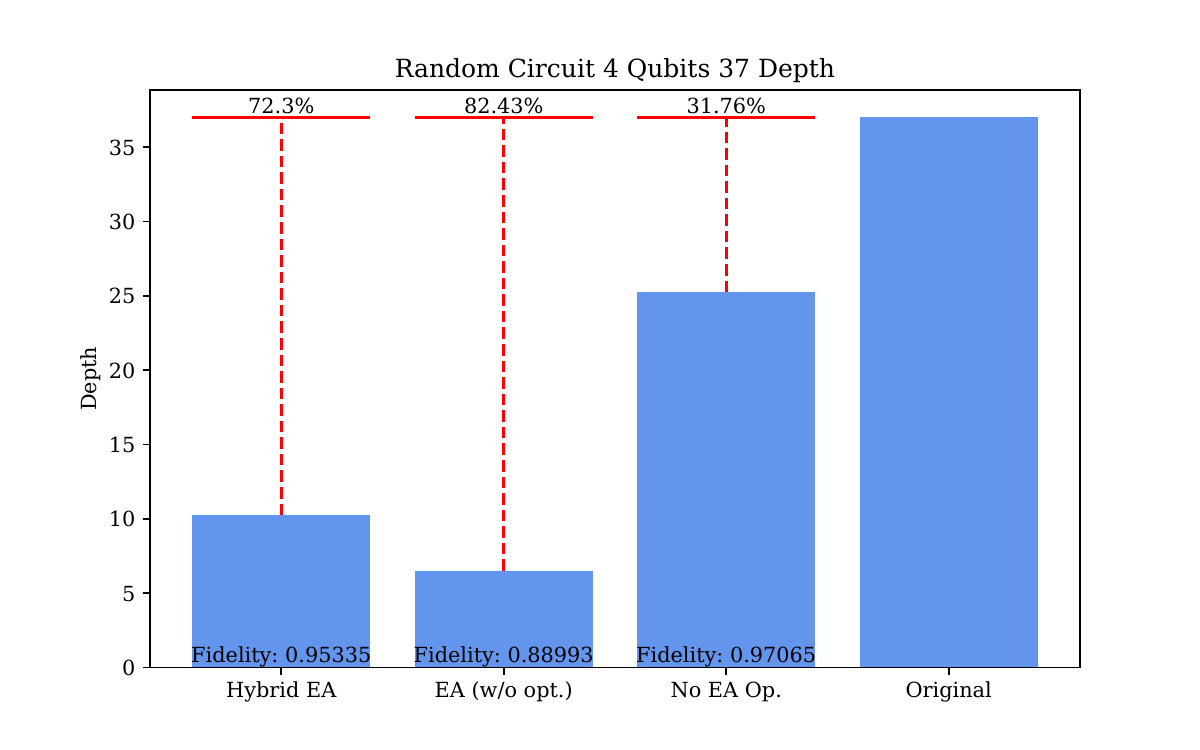}
        \caption{4 qubits and 37 depth.}
        \label{fig:results_qc_from_scratch_c}
    \end{subfigure}
    \begin{subfigure}[t]{0.45\textwidth}
        \centering
        \includegraphics[scale=0.3]{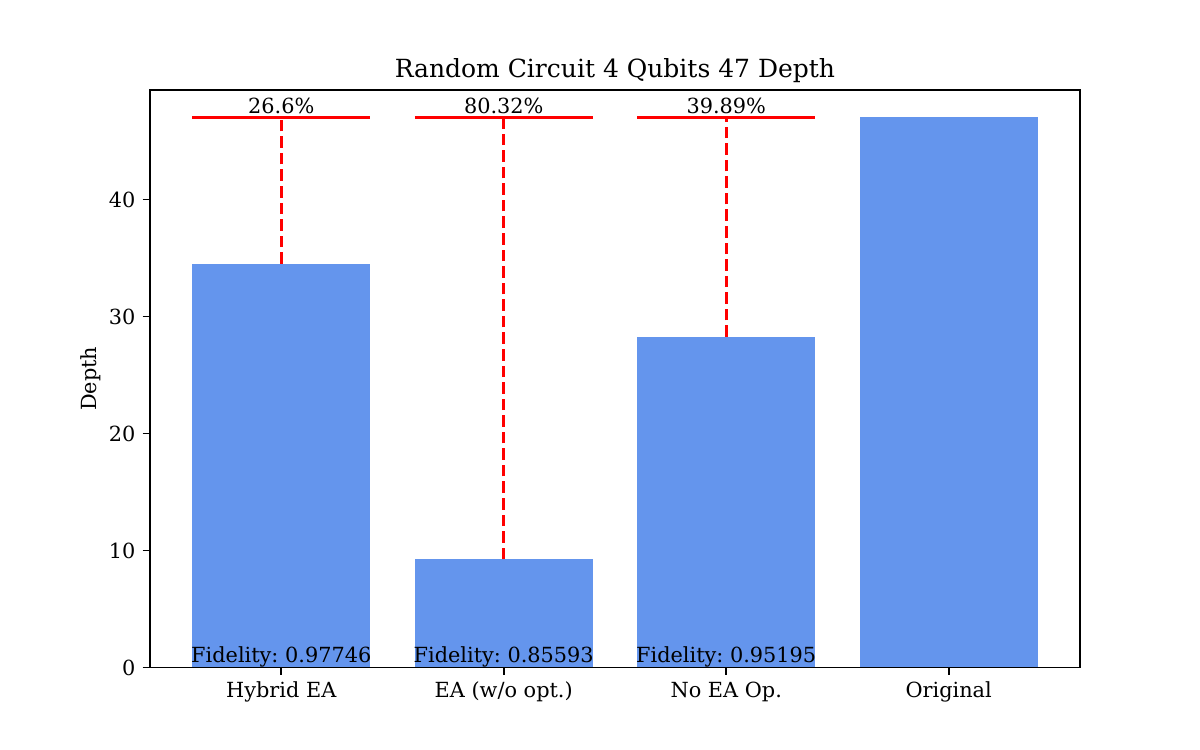}
        \caption{4 qubits and 47 depth.}
        \label{fig:results_qc_from_scratch_d}
    \end{subfigure}
    \caption{Results of circuits with 4 qubits where the circuits are initialized randomly, i.e., created from scratch. Numbers on the top specify the achieved reduction in depth.}
    \label{fig:results_qc_from_scratch}
\end{figure*}

\section{Results}\label{sec:results}
We present and discuss the experimental results in this section. We begin with the experiments in which circuits are constructed from the ground up, i.e., the individual's solution is initialized randomly with a random depth, so the algorithm does not start with an empty circuit. We then discuss the results when the target circuit is used to initialize individuals which are then subsequently optimized. 

\subsection{Circuit Construction from Scratch}
Each circuit was created randomly and the algorithms were run for 4 different seeds. The mean results for these experiments are shown in Fig. \ref{fig:results_qc_from_scratch}. Note that fidelities are rounded in the plot (5 decimals) as well as in the following text (2 decimals). In the first experiment (Fig. \ref{fig:results_qc_from_scratch_a}) with a circuit of depth 20, the hybrid EA achieves a depth reduction of 80\% with a fidelity of roughly 0.98 whereas the EA without parameter optimization is able to reduce the depth by 82.5\%, with a fidelity of 0.97. The approach where the crossover and mutations are turned off also achieves a fidelity of 0.95; it is not able to find circuits that reduce the depth as good as the algorithms containing EA operations, however, but it is still able to reduce the depth by 53.75\%.
In the next experiment with a larger circuit (Fig. \ref{fig:results_qc_from_scratch_b}), a similar pattern emerges. Both approaches with a parameter optimization routine achieve a high fidelity of 0.96, but are not able to reduce the depth as much as the EA without parameter optimization, which has a fidelity of 0.92. The EA achieves a depth reduction of 75\%, the hybrid of roughly 53.12\%, whereas the third approach with no EA operations only of 14.84\%.
With a slightly larger circuit with depth of 37, the hybrid and regular EA achieve a fidelity of 0.95 and 0.89 respectively and are able to reduce the depth by 72.3\% and 82.43\%; whereas the baseline achieves a higher fidelity (0.97), it reduces the depth by 31.76\%, i.e., in terms of depth minimization, the EA vastly outperforms the baseline, as can be seen in Fig. \ref{fig:results_qc_from_scratch_c}. 
Figure \ref{fig:results_qc_from_scratch_d} depicts the results of the experiment with the circuit of depth 47. The hybrid EA and the baseline achieve the best fidelity with 0.98 and 0.95, respectively, while the EA only achieves 0.86. However, in terms of depth, the EA achieves the best results with an improvement of 80.32\% whereas the hybrid 26.6\% and the baseline 39.89\%.

\begin{figure*}
    \centering
    \begin{subfigure}[t]{0.45\textwidth}
        \centering
        \includegraphics[scale=0.3]{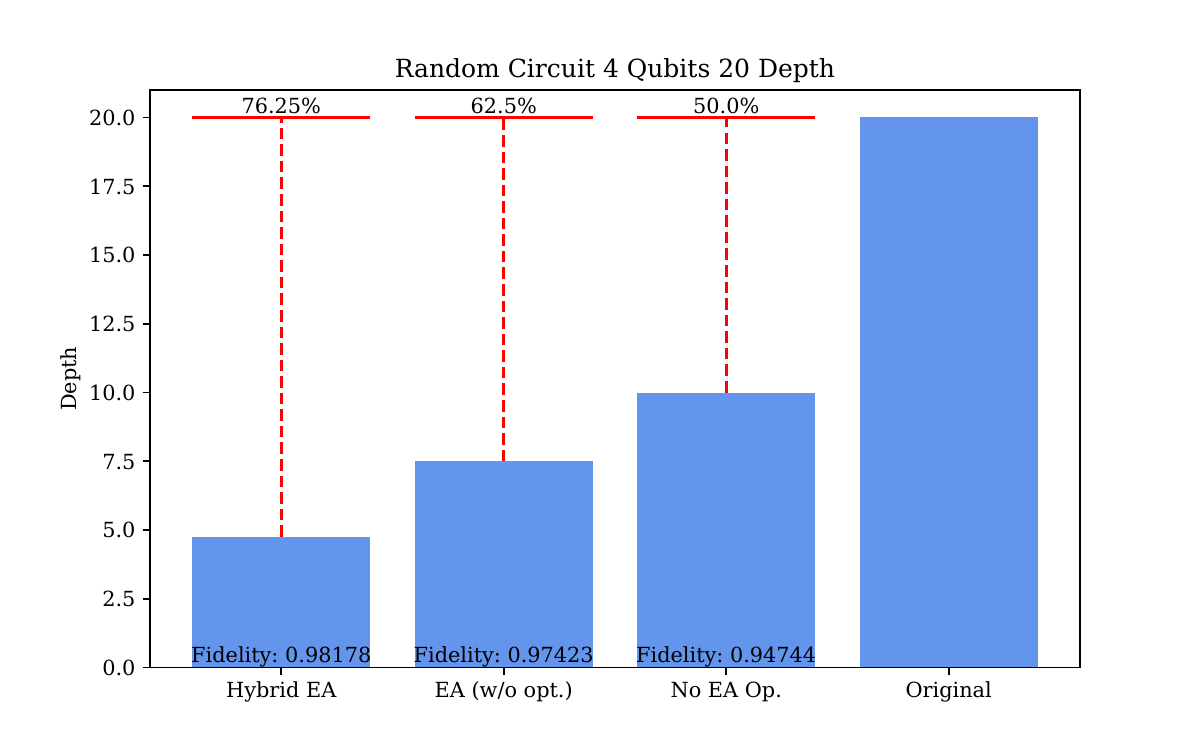}
    \end{subfigure}
    \begin{subfigure}[t]{0.45\textwidth}
        \centering
        \includegraphics[scale=0.3]{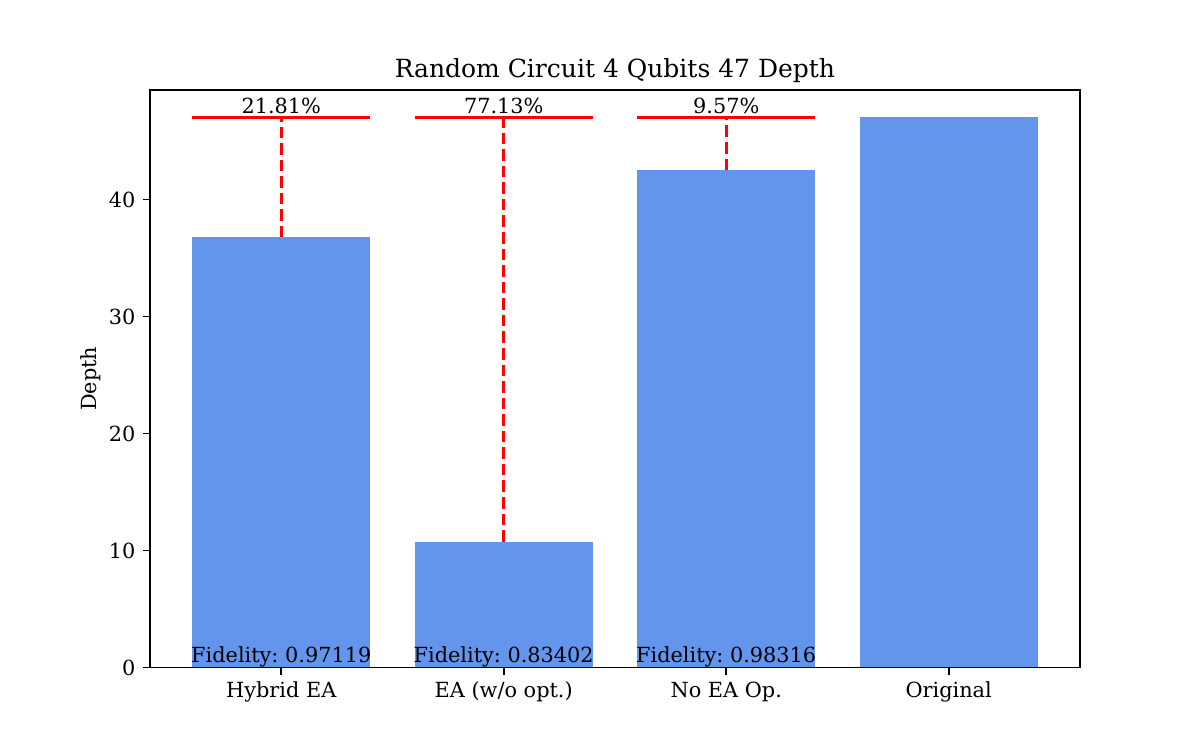}
    \end{subfigure}
    \caption{Results of experiments without the solution optimization subroutine}
    \label{fig:results_no_solution_optimization}
\end{figure*}

Lastly, Figure \ref{fig:results_no_solution_optimization} shows the results of the circuits with a depth of 20 and 47 where the solution optimization subroutine is not applied. 

\begin{figure*}
    \centering
    \begin{subfigure}[t]{0.45\textwidth}
        \centering
        \includegraphics[scale=0.3]{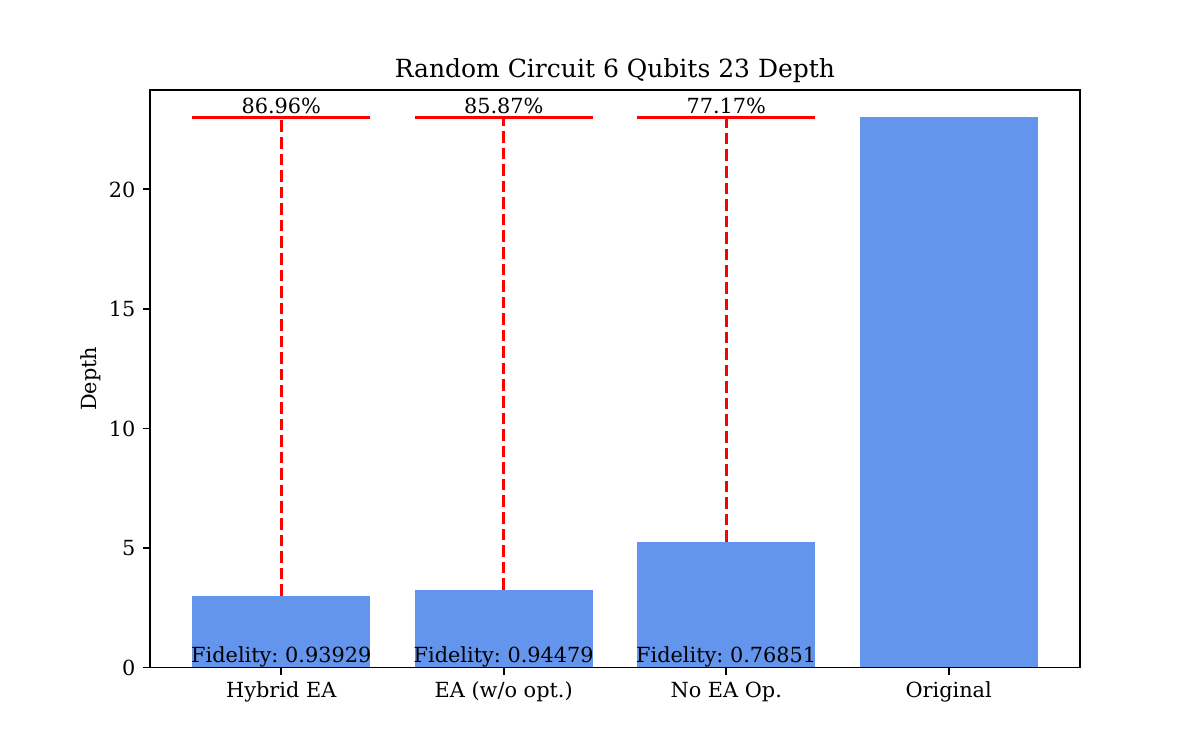}
    \end{subfigure}
    \begin{subfigure}[t]{0.45\textwidth}
        \centering
        \includegraphics[scale=0.3]{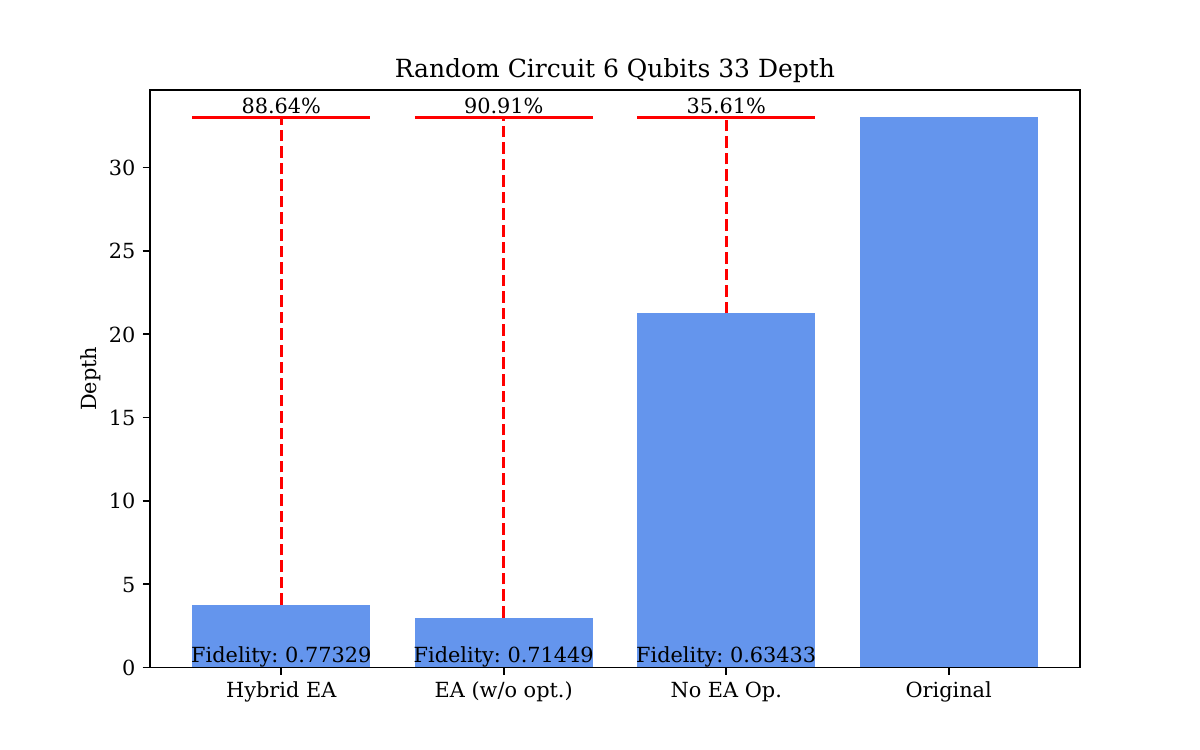}
    \end{subfigure}
    \caption{Results of the experiments with circuits containing 6 qubits.}
    \label{fig:results_qc_from_scratch_6_qubits}
\end{figure*}

\begin{figure*}
    \centering
    \begin{subfigure}[t]{0.45\textwidth}
        \centering
        \includegraphics[scale=0.3]{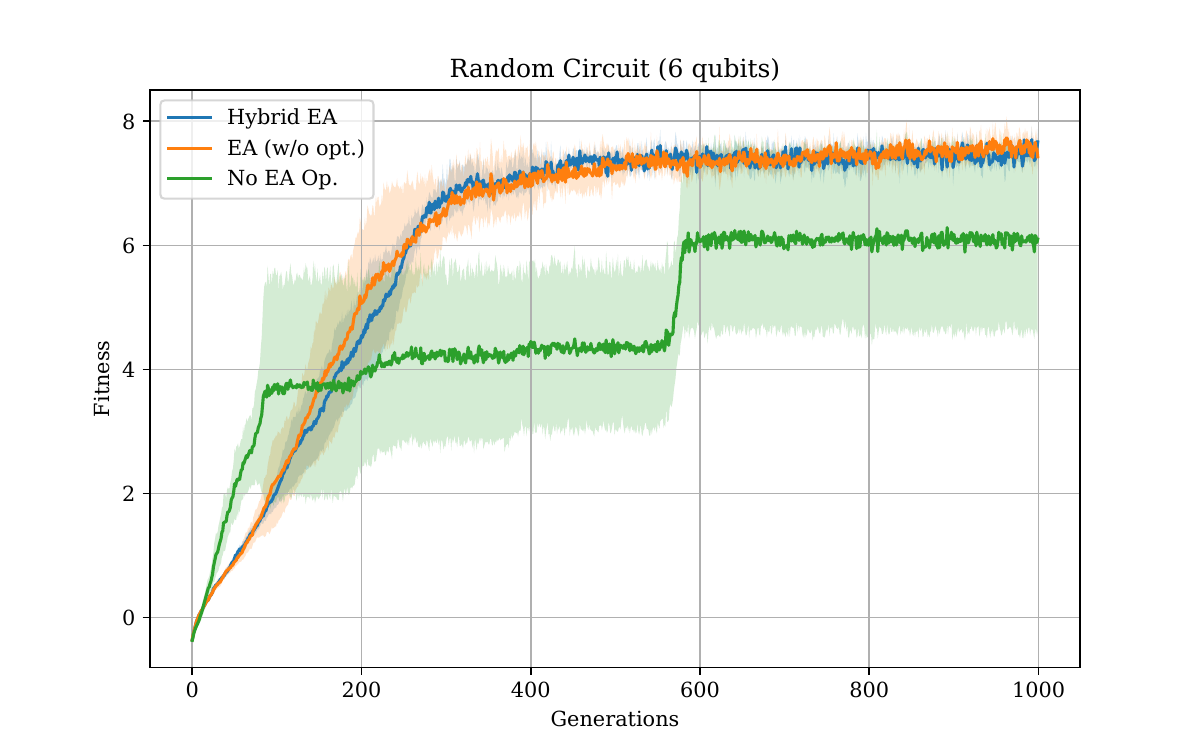}
        \caption{Mean fitness per generation}
    \end{subfigure}
    \begin{subfigure}[t]{0.45\textwidth}
        \centering
        \includegraphics[scale=0.3]{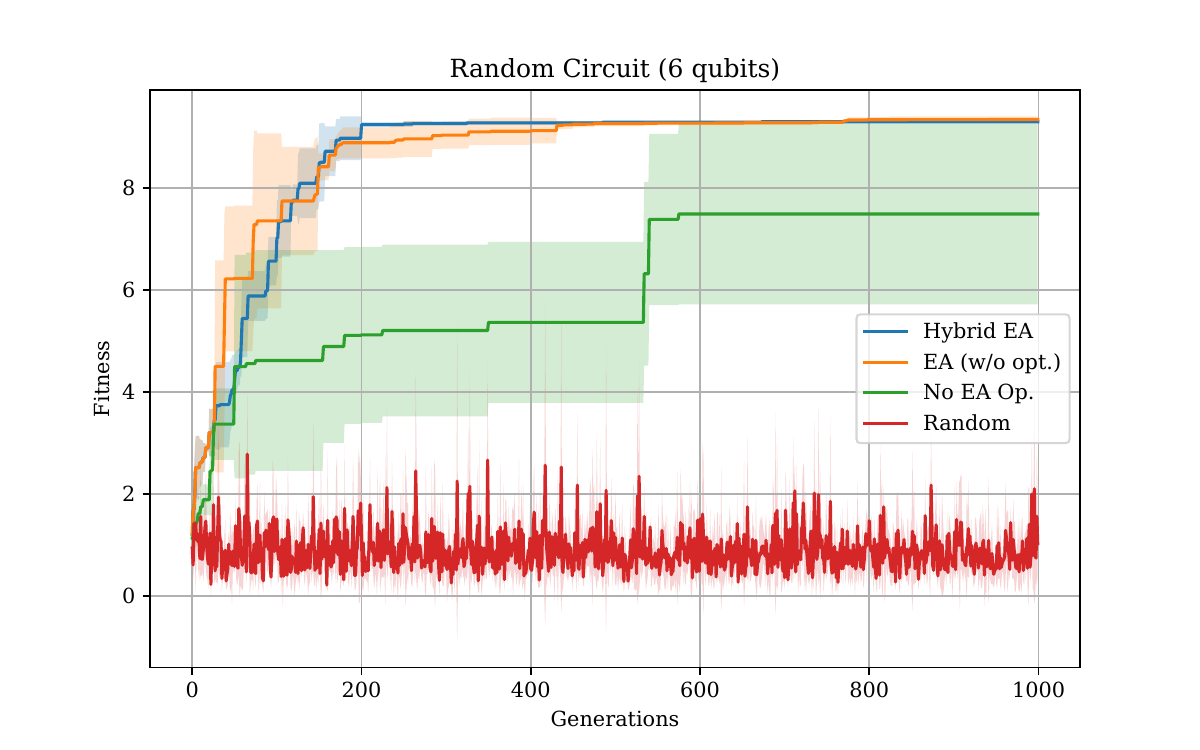}
        \caption{Best fitness per generation}
    \end{subfigure}
    \caption{Fitness over the generations for experiments with circuit with 6 Qubits and depth of 23. Plot shows aggregated results and std.}
    \label{fig:fitness_per_gen_qc_from_scarth_6_qubits}
\end{figure*}

In the smaller circuit, all approaches achieve an identical fidelity but a decreased improvement in terms of depth if the solution optimization routine is turned off. While the difference in the algorithms that use a parameter optimization subroutine is relatively low (roughly 4\%), whereas for the regular EA the difference is 20\%.
In the largest circuit with a depth 47 the difference between the two approaches is also visible. The fidelity remains similar; without the solution optimization subroutine the EAs fidelity slightly decreases from 0.98 to 0.97 and from 0.86 to 0.83 for the hybrid and regular EA respectively. The baseline, however, achieves a slightly better fidelity as it increases from 0.95 to 0.98. However, depth improvement decreases for all three approaches from 26.6\% to 21.81\%, 80.32\% to 77.13\% and 39.89\% to 9.57\% for the hybrid, regular EA and baseline respectively.
While the improvement is already clear, the subroutine could be extended to be exhaustive, i.e., optimize each circuit even further to make it as compact as possible, although this would increase the computational cost. Ultimately, more investigation would be required to determine how to balance the procedure to obtain optimal results.
Figure \ref{fig:results_qc_from_scratch_6_qubits} shows the results for the experiments with circuits containing 6 qubits. In the circuit with a depth of 23, both EAs outperform the baseline in both metrics, achieving a depth reduction 86.96\% and 85.87\% for the hybrid and regular EA respectively; both result in a fidelity of 0.94. The baseline has a fidelity of 0.77 and a depth improvement of 77.17\%. In the second circuit, the EAs again outperform the baseline on both metrics, achieving again a depth improvement of over 88.64\% and 90.91\% while the fidelity drops to 0.77 and 0.71 for the hybrid and regular EA respectively. The baseline is able to reduce the depth by 35.61\% with a fidelity of only 0.63. The fitness over the generations is depicted in Figure \ref{fig:fitness_per_gen_qc_from_scarth_6_qubits}. Both EAs perform similarly well; however, the hybrid approach converges earlier and does not require as many generations.

\subsection{Circuit Optimization}
\begin{figure*}
    \centering
    \begin{subfigure}[t]{0.45\textwidth}
    \centering
        \includegraphics[scale=0.3]{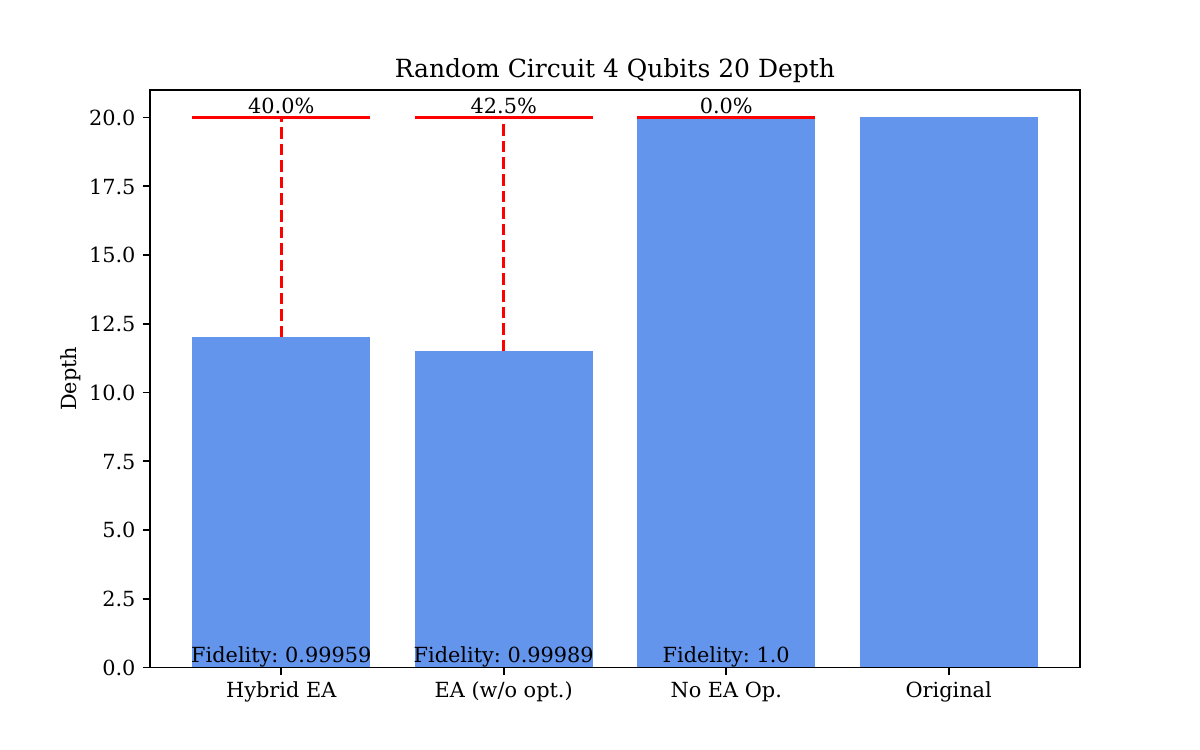}
        \caption{4 Qubits and depth of 20.}
        \label{fig:results_qc_optimization_a}
    \end{subfigure}
    \begin{subfigure}[t]{0.45\textwidth}
    \centering
        \includegraphics[scale=0.3]{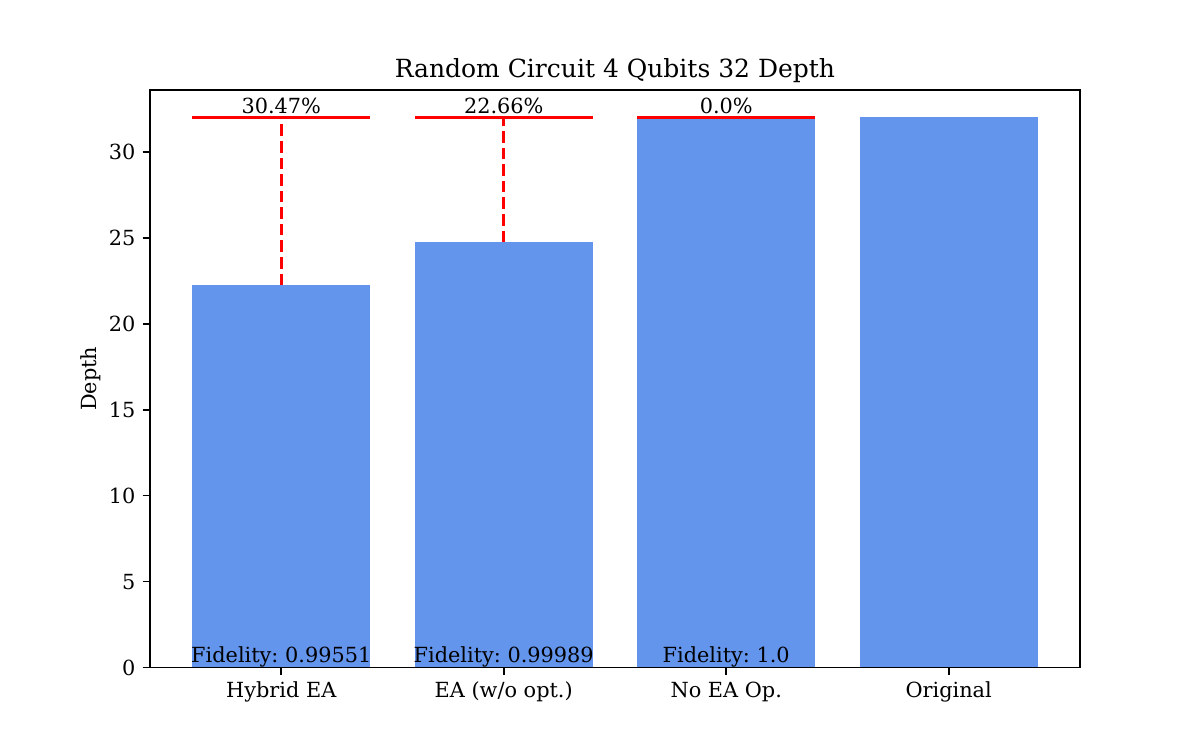}
        \caption{4 Qubits and depth of 32.}
        \label{fig:results_qc_optimization_b}
    \end{subfigure}
    \begin{subfigure}[t]{0.45\textwidth}
    \centering
        \includegraphics[scale=0.3]{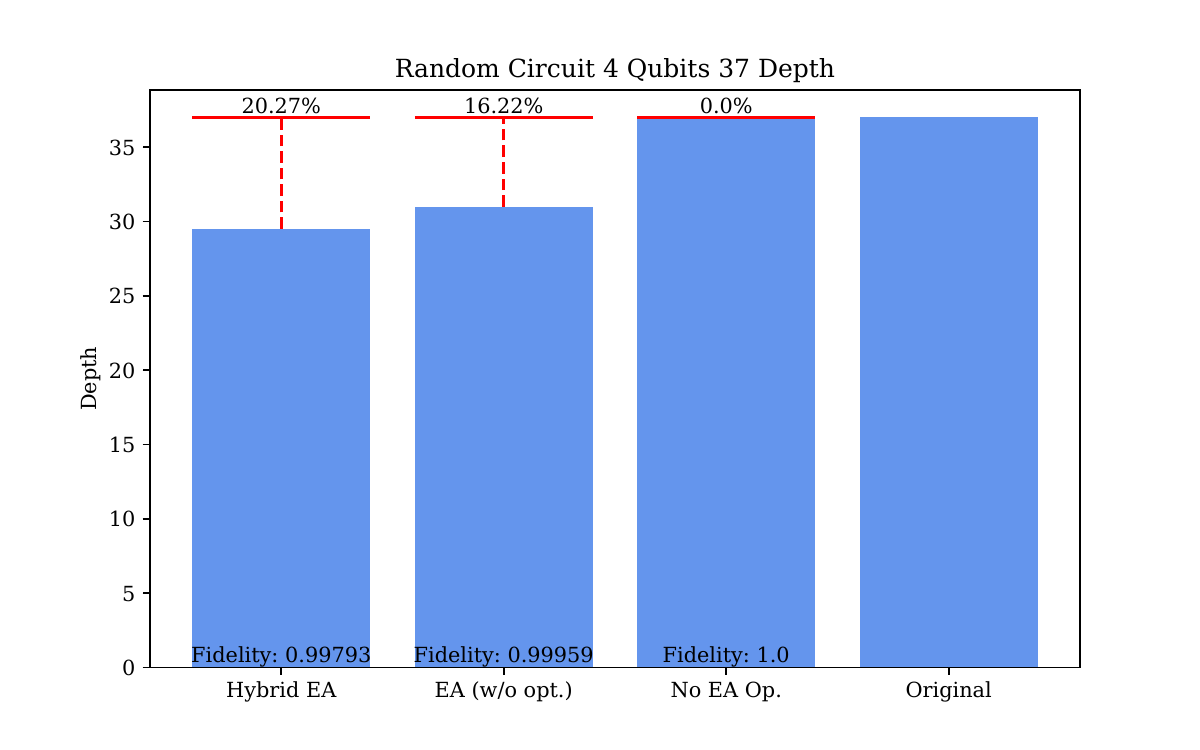}
        \caption{4 Qubits and depth of 37.}
        \label{fig:results_qc_optimization_c}
    \end{subfigure}
    \begin{subfigure}[t]{0.45\textwidth}
    \centering
        \includegraphics[scale=0.3]{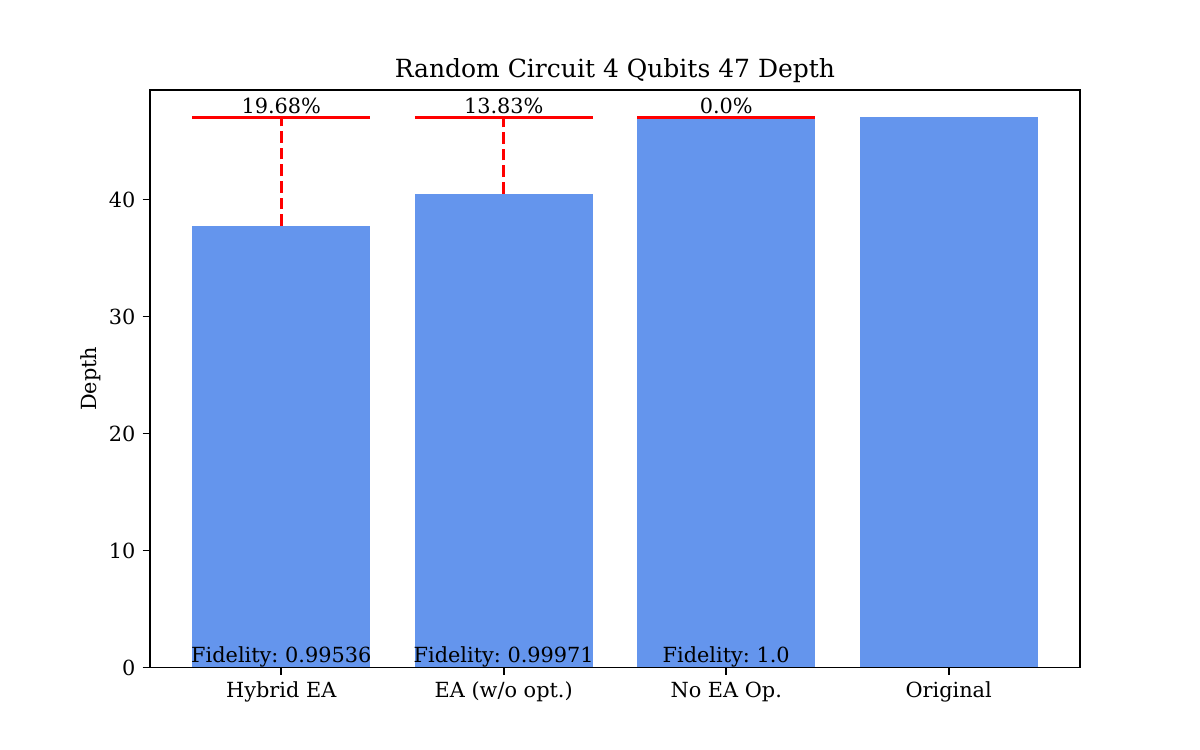}
        \caption{4 Qubits and depth of 47.}
         \label{fig:results_qc_optimization_d}
    \end{subfigure}

    \caption{Results of the circuit optimization experiments, i.e., population is initialized with the target as solution.}
    \label{fig:results_qc_optimization}
\end{figure*}

In these experiments, the population contains individuals whose solutions are initialized with the target circuit, and the aim is to reduce the depth while maintaining a high fidelity. Results for these experiments are depicted in Fig. \ref{fig:results_qc_optimization}.
With the smallest circuit (Fig. \ref{fig:results_qc_optimization_a}), all approaches achieve a fidelity over 0.99, and both evolutionary approaches achieve a similar depth reduction (40\% and 42.5\%). For the approach with no EA operations, the fidelity remains stable at 1.0 with no reduction of the depth. In the circuits with depth 32 and 37, the EA approaches achieve a fidelity roughly 0.99 and depth improvements of 30.47\% and 20.27\% for the hybrid and 22.66\% and 16.22\% for the EA.
For the largest circuit, results are similar. The EA approaches maintain a high fidelity of roughly 0.99 while achieving a depth reduction of 19.68\% and 13.83\% for the hybrid and EA respectively. As with the smaller circuits, the approach with no EA operations remains at a fidelity of 1 and no depth reduction.
As the population (and new random individuals introduced into the population that are not created by crossover) is initialized with the target as solution, its fidelity value will start at 1, signifying identical states. The aim is then to adjust the circuit in such a way to reduce the depth in order to increase the fitness value. Using this approach, the algorithms are able to retain a high fidelity.

\subsection{Discussion}\label{sec:discussion}
To summarize, we ran experiments approaching the problem of circuit construction and optimization from two directions. In the first, circuits are constructed from the ground up (i.e., initialized randomly) while in the second a target is given as the initialization point. While the latter approach is able to achieve high fidelity values in all experiments, the first approach is able to reduce the circuit's depth significantly better. As both objectives are important for this use case, it is not necessarily straightforward to determine which approach is better, although a high fidelity is a crucial aspect. Adjusting the weights in the fitness function may allow for more control of the balance between the different components in the fitness. 
A further aspect of our analysis is the difference between the hybrid and regular EA approaches. In the majority of experiments where circuits are created from scratch, the fidelity of the EA without parameter optimization is worse than approaches that have this subroutine. However, the regular EA achieves the best improvement in terms of depth in most experiments. Thus, the question of what is more important arises, and it is also noteworthy that the hybrid is computationally more expensive, as it has to run a parameter optimization for a subset of individuals. All algorithms also contain a solution optimization subroutine, which also affects the performance. Overall, the impact of this technique is positive; however, in some instances the fidelity might get better without it while the depth improvement decreases, which again comes back to the questions of balance and what ultimately should be achieved.

\section{Conclusion}\label{sec:conclusion}
Constructing novel circuits or optimizing existing ones that perform some desired task while also fulfilling specified constraints is an important research avenue at the current stage of quantum computing. Automating this process through machine learning algorithms like RL or optimization algorithms such as EAs is currently being heavily investigated and studied by the community. In this paper, we proposed and evaluated a hybrid evolutionary approach that aims to construct quantum circuits that maximize their fidelity to some target state while minimizing the circuits' depth. The hybrid algorithm makes use of a parameter optimization subroutine that adjusts the angles of rotation gates in order to maximize the fidelity to the target. We compared this approach to a regular EA and a baseline without evolutionary operations. 
Additionally, we approached the problem from two sides: (1) constructing circuits from the ground up and (2) starting with a population with the target as initial solution. The results indicate that for high fidelity  the latter approach is better suitable while the former if the depth reduction is more important. The same is true for the hybrid and regular EAs. While one performed better in terms of fidelity, the other achieved a better depth improvement. Ultimately, further investigation is required to ensure an even balance between the two objectives in practical relevant settings.

\begin{acks}
Sponsored in part by the Bavarian Ministry of Economic Affairs, Regional Development and Energy as part of the 6GQT project. The research is also part of the Munich Quantum Valley, which is supported by the Bavarian state government with funds from the Hightech Agenda Bayern Plus.
\end{acks}

\bibliographystyle{ACM-Reference-Format}
\bibliography{bibliography}

\end{document}